\documentclass{ws-ijmpe}
\begin{document}

\markboth{Trivedi, Srivastava, Negi and Mehrotra}{Shell model description of $^{102-108}$Sn isotopes}

\catchline{}{}{}{}{}
%%%%%%%%%%%%%%%%%%%%%%%%%%%%%%%%%%%%%%%%%%%%%%%%%%%%%%%%%%%%%%%%%%%%

\title{SHELL MODEL DESCRIPTION OF $^{102-108}$Sn ISOTOPES}

\author{T.~TRIVEDI}
\address{Tata Institute of Fundamental Research, Mumbai- 400005, India and \\ 
Department of Physics, University of Allahabad, Allahabad-211002, India}
\author{ P.C.~SRIVASTAVA}
\address{Instituto de Ciencias Nucleares, Universidad Nacional Aut\'onoma de
M\'exico, Apartado Postal 70-543, 04510 M\'exico, D.F., M\'exico}
\author{D. NEGI}
\address{Inter University Accelerator Centre, New Delhi- 1160067, India}
\author{ I.~MEHROTRA}
\address{Department of Physics, University of Allahabad, Allahabad-211002, India}
%\author{T.~TRIVEDI$^{1,2}$, P.C.~SRIVASTAVA$^{3}$, D. NEGI$^{4}$ and I.~MEHROTRA$^{1}$}

\maketitle

\begin{history}
%\received{(received date)}
%\revised{(revised date)}
%\accepted{(Day Month Year)}
%\comby{(xxxxxxxxxx)}
\end{history}

\begin{abstract}
       We have performed shell model calculations for neutron deficient even $^{102-108}$Sn and odd $^{103-107}$Sn isotopes in $sdg_{7/2}h_{11/2}$ model space using two different interactions. The first set of interaction is due to Brown {\it et al.} and second is due to Hoska {\it et al.} The calculations have been performed using doubly magic $^{100}$Sn as core and valence neutrons are distributed over the single particle orbits 1$g_{7/2}$, 2$d_{5/2}$, 2$d_{3/2}$, 3$s_{1/2}$ and 1$h_{11/2}$. In more recent experimental work for $^{101}$Sn [Phys. Rev. Lett. {\bf 105} (2010) 162502], the g.s. is predicted as  5/2$^+$ with excited 7/2$^+$ at 172 keV. We have also performed another two set of calculations by taking difference in single particle energies of 2$d_{5/2}$ and 1$g_{7/2}$ orbitals by 172 keV. The present state-of-the-art shell model calculations predict fair agreement with the experimental data. These calculations serve as a test of nuclear shell model in the region far from stability  for unstable Sn isotopes near the doubly magic $^{100}$Sn core.
\end{abstract}

\maketitle

\section{\label{Intro}Introduction}

The structure of neutron deficient nuclei far from $\beta$-stability is currently at the focus of nuclear structure  due to development of the radioactive ion beam facilities, high efficiency gamma detector array and state-of-the-art shell model calculations using new effective interactions. Further, nuclei near the N$\simeq$Z, heaviest doubly magic core are more interesting because they exhibit many new structural phenomenon such as shell evolution \cite {Ots10}, change of collective properties \cite{Fed79}, band termination \cite{Wad96} and magnetic rotation \cite{Gad97} $etc.$

In recent years rich variety of experimental data is now available for neutron deficient Sn isotopes by various experimental studies. Faestermann {\it et al.} first time measured the half-life,  $\beta$- and  $\gamma$-spectrum of $^{102}$Sn at GSI \cite{Fae02}. Exited states for $^{103}$Sn using EUROBALL and ancillary detectors
was reported in \cite{Fah01}. Structure of high spin states in $^{104}$Sn using reaction $^{58}$Ni($^{50}$Cr,2$p$2$n$) at 200 and 205 MeV beam energies were reported by G\'orska {\it et al.} \cite{Gor98}. Gadea  {\it et al.} \cite{Gad97} investigated structure of $^{105}$Sn through $^{50}$Cr($^{58}$Ni,2$p$$n$) reaction at energies of 210 MeV, they also found dipole band similar to the Pb region. Coexistence of collective and quasiparticle structure in $^{106}$Sn and $^{108}$Sn have been reported by Juutinen {\it et al.} \cite{Juu97}. For $^{108}$Sn smooth band termination is also reported in \cite{Wad96}. In more recent experiment Banu {\it et al.} studied
$^{108}$Sn in inverse kinematics by intermediate-energy Coulomb excitation using the RISING setup at GSI \cite {Ban05}. Vaman {\it et al.} studied  even-mass $^{106-112}$Sn at NSCL from intermediate-energy Coulomb excitations \cite {Vam07}. Recently Ressler {\it et al.} reported $\beta$ decay studies of $^{107,109}$Sb to establish low-lying states in $^{107,109}$Sn \cite {Res02}.

From theoretical point of view the Sn isotopes are interesting to study, as they are unique testing ground for nuclear structure calculation. Previously shell model calculation for neutron-deficient Sn isotopes have been performed by Engeland {\it et al.} \cite {Eng95}, Covello {\it et al.} \cite {Cov97}  and Schubert {\it et al.} \cite {Sch95} taking $^{100}$Sn as core, almost more than one decade back.
However, recently shell model calculations for $^{106-109}$Sn isotopes using CD-Bonn and Nijmegen1 interactions are reported by Dikman \cite{Dik09}. In the present work, large - scale shell model calculation have been performed for neutron deficient even $^{102-108}$Sn and odd $^{103-107}$Sn isotopes for more recent accumulated experimental data.
The calculation have been performed in  $sdg_{7/2}h_{11/2}$ model space.
The aim of present work is to test the effective interactions for these lighter Sn isotopes near the N$\simeq$Z line where pairing correlation are more important.

\begin{table}[http]
\begin{center}
\caption{Single particle energies used in different shell model approaches.}
\label{spe}
%\resizebox{7.5cm}{3.5cm}{
\begin{tabular}{rcrcrcrc}
\hline

 orbital & ~~sn100pn &snet~~ & sn100pn(mod.) &snet(mod.)\\
 ~~~ nlj~~~~ &&Energy(MeV)~~~~	\\
\hline
2d$_{5/2}$ &~~ -10.6089 & -10.10 &~~ -10.6089 & -10.322 \\
1g$_{7/2}$ &~~ -10.2893 & -10.15 &~~ -10.4369 & -10.15 \\
3s$_{1/2}$ &~~ -8.7167 & -8.40  &~~ -8.7167 & -8.40  \\
2d$_{3/2}$ &~~ -8.6944 & -8.09 &~~ -8.6944 & -8.09  \\
1h$_{11/2}$ &~~ -8.8152 & -7.85 &~~ -8.8152 & -7.85  \\

\hline

\end{tabular}
\end{center}
\end{table}

\section{\label{Exp}Details of Calculation}

Shell model calculation of the low-lying states for mass A = 102-108, Sn isotopes have been performed in $sdg_{7/2}h_{11/2}$ valence space taking $^{100}$Sn as core. Valence neutrons are distributed over the single particle orbits 1$g_{7/2}$, 2$d_{5/2}$, 2$d_{3/2}$, 3$s_{1/2}$ and 1$h_{11/2}$.
In the present calculations, we have used two sets of interactions. The first set of interaction named sn100pn was obtained from a realistic interaction developed by Brown {\it et al.,} \cite {Bro05} starting with G matrix derived from CD Bonn nucleon nucleon interaction \cite{Mac96}.
This interaction has three parts, proton-proton (pp), neutron-neutron (nn) and proton-neutron (pn). Apart from Coulomb interaction was added to the interaction between protons. The second set of interaction named snet was obtained from bare G matrix by Hoska {\it et al.} \cite {Hos85} based on renormalized Paris potential for $N$ = 82 nuclei \cite {Bro94}.  The single particle energies for these two interactions are given in Table 1.
In more recent work  for $^{101}$Sn, the g.s. is predicted as  5/2$^+$ with excited 7/2$^+$ at 172 keV \cite{Darby10}. We have also performed another two set of calculations by taking difference in single particle energies of 2$d_{5/2}$ and 1$g_{7/2}$ orbitals by 172 keV. These two modified results are reported as sn100pn(mod.) and snet(mod.).
The calculation have been performed using code Nushell \cite {Bro07} without any truncation except in case of $^{108}$Sn where we have put maximum two particles in 1$h_{11/2}$ orbital.

\section{\label{Results} Results and Discussion}
\subsection{Excitation energies of even Sn isotopes}

\begin{figure}
\begin{center}
\includegraphics[width=10.5cm]{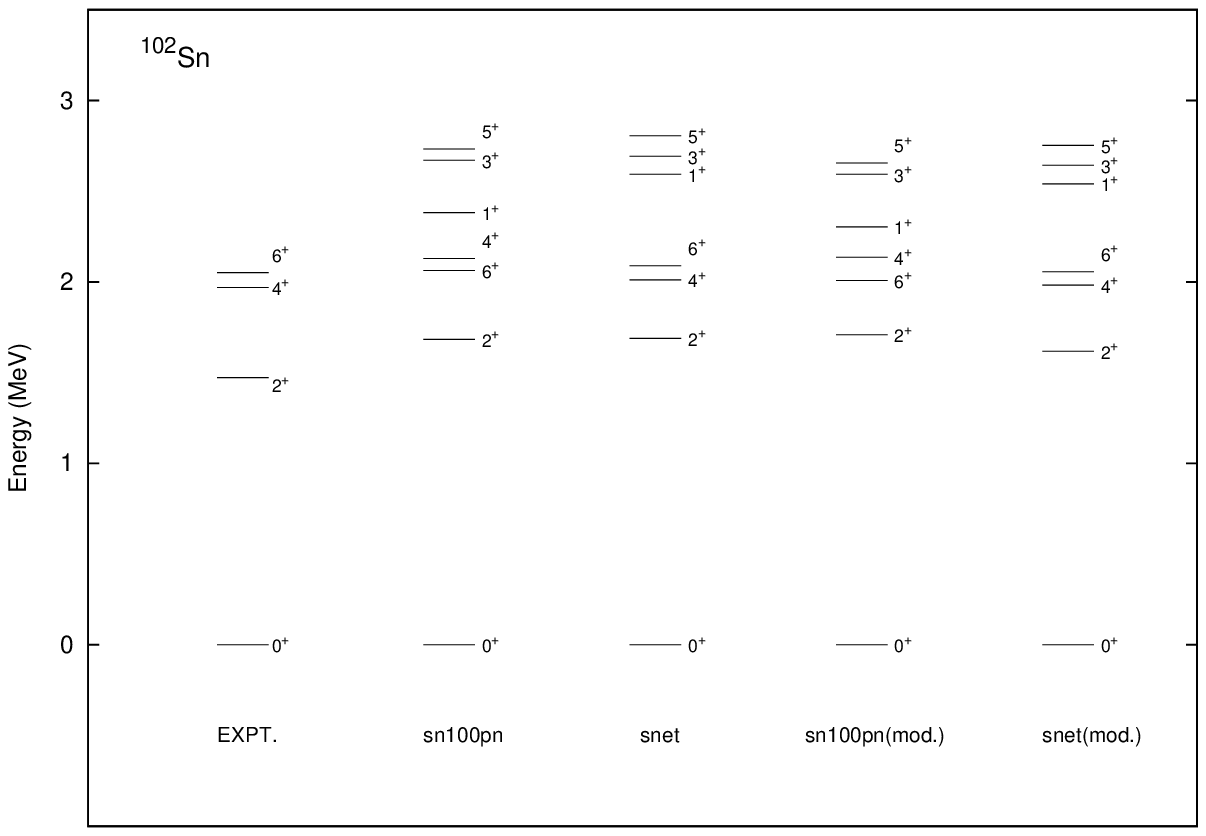}% Here is how to import EPS art
\caption{\label{102Sn} Experimental \protect\cite {Lip98} and calculated energy levels of $^{102}$Sn with sn100 and snet interactions.}
\end{center}
\end{figure}

In Table 2, we have tabulated all the calculated and experimentally known levels for $^{102}$Sn and corresponding  levels are also shown in Fig.\ \ref{102Sn}. Experimentally, only few excited states with J$^\pi$= 2$^+$, 4$^+$ and 6$^+$ are known. The ground state 0$^+$ is well predicted by both the interactions and it is found that calculated 2$^+$ state lies at 211 and 217 keV above the experimental values for sn100pn and snet interactions.
However, with modified single particle energies\cite{Darby10} in snet(mod.) interaction, the calculated 2$^+$ states is well reproduced with an energy difference of 146 keV, whereas, from sn100pn(mod.) interaction change is not much pronounced. The 4$^+$ state is well predicted by snet interaction but in case of sn100pn interaction it lies above the 6$^+$. The 6$^+$ is predicted at 2063 and 2089 keV by sn100pn and snet interaction, respectively.
\begin{table}
\begin{center}
\caption{Experimental \protect\cite{Lip98} and theoretical low-lying states of $^{102}$Sn. Energies are in MeV.}
\label{102sn}
\resizebox{12.7cm}{!}{
\begin{tabular}{rcrcrcrcrcrc}
\hline

 J$^\pi$ & ~~Exp. &~~ J$^\pi$ &~~ sn100pn &~~  J$^\pi$ &~~ snet~~ &~~ J$^\pi$ &~~ sn100pn(mod.) &~~  J$^\pi$ &~~ snet(mod.)~~\\
\hline
(0$^+$) &~~ 0.000 &~~ 0$^+$  &~~ 0.000 &~~ 0$^+$ &~~ 0.000 &~~ 0$^+$  &~~ 0.000 &~~ 0$^+$ &~~ 0.000  \\
(2$^+$) &~~ 1.472 &~~ 2$^+$  &~~ 1.683 &~~ 2$^+$ &~~ 1.689 &~~ 2$^+$  &~~ 1.709 &~~ 2$^+$ &~~ 1.618 \\
(4$^+$) &~~ 1.969 &~~ 6$^+$  &~~ 2.063 &~~ 4$^+$ &~~ 2.011  &~~ 6$^+$  &~~ 2.007 &~~ 4$^+$ &~~ 1.982\\
(6$^+$) &~~ 2.051 &~~ 4$^+$  &~~ 2.129 &~~ 6$^+$ &~~ 2.089 &~~ 4$^+$  &~~ 2.136 &~~ 6$^+$ &~~ 2.056 \\
   -   &~~  -     &~~ 0$^+$  &~~ 2.301 &~~ 2$^+$  &~~ 2.484  &~~ 0$^+$  &~~ 2.178 &~~ 2$^+$  &~~ 2.420\\
   -   &~~  -    &~~ 1$^+$  &~~ 2.381 &~~ 0$^+$  &~~ 2.565 &~~ 1$^+$  &~~ 2.303 &~~ 0$^+$  &~~ 2.487 \\
   -   &~~  -    &~~ 2$^+$  &~~ 2.396 &~~ 1$^+$  &~~ 2.593 &~~ 2$^+$  &~~ 2.305 &~~ 4$^+$  &~~ 2.518 \\
   -   &~~   -   &~~ 6$^+$  &~~ 2.582 &~~ 4$^+$  &~~ 2.666 &~~ 4$^+$  &~~ 2.583 &~~ 1$^+$  &~~ 2.540\\
   -   &~~   -   &~~ 4$^+$  &~~ 2.617 &~~ 6$^+$  &~~ 2.680 &~~ 3$^+$  &~~ 2.593 &~~ 3$^+$  &~~ 2.643\\
   -   &~~   -   &~~ 3$^+$  &~~ 2.670 &~~ 3$^+$  &~~ 2.692  &~~ 6$^+$  &~~ 2.629 &~~ 5$^+$  &~~ 2.753\\
   -   &~~   -   &~~ 5$^+$  &~~ 2.733 &~~ 5$^+$  &~~ 2.805 &~~ 5$^+$  &~~ 2.655 &~~ 6$^+$  &~~ 2.831\\
\hline

\end{tabular}}
\end{center}
\end{table}

\begin{table}
\begin{center}
\caption{
Experimental \protect\cite {Gor98} and theoretical low-lying states (up to 4.5 MeV) of $^{104}$Sn. Energies are in MeV.}
\label{104Sn}
\resizebox{12.7cm}{!}{
\begin{tabular}{rcrcrcrcrcrc}
\hline

 J$^\pi$ & ~~Exp. &~~ J$^\pi$ &~~ sn100pn &~~  J$^\pi$ &~~ snet~~ &~~ J$^\pi$ &~~ sn100pn(mod.) &~~  J$^\pi$ &~~ snet(mod.)~~\\
\hline
(0$^+$) &~~ 0.000 &~~ 0$^+$  &~~ 0.000 &~~ 0$^+$ &~~ 0.000 &~~ 0$^+$  &~~ 0.000 &~~ 0$^+$ &~~ 0.000 \\
(2$^+$) &~~ 1.260 &~~ 2$^+$  &~~ 1.494 &~~ 2$^+$ &~~ 1.597 &~~ 2$^+$  &~~ 1.467 &~~ 2$^+$ &~~ 1.563  \\
(4$^+$) &~~ 1.943 &~~ 4$^+$  &~~ 2.111 &~~ 4$^+$ &~~ 2.098 &~~ 4$^+$  &~~ 2.067 &~~ 4$^+$ &~~ 2.070  \\
(6$^+$) &~~ 2.257 &~~ 6$^+$  &~~ 2.282 &~~ 6$^+$ &~~ 2.188 &~~ 6$^+$  &~~ 2.246 &~~ 6$^+$ &~~ 2.162 \\
(8$^+$) &~~ 3.440 &~~ 0$^+$  &~~ 2.358 &~~ 0$^+$ &~~ 2.337 &~~ 0$^+$  &~~ 2.358 &~~ 0$^+$ &~~ 2.201 \\
(10$^+$) &~~ 3.981 &~~ 2$^+$  &~~ 2.537 &~~ 2$^+$ &~~ 2.534  &~~ 2$^+$  &~~ 2.476 &~~ 4$^+$ &~~ 2.442 \\
   -   &~~  -     &~~ 1$^+$  &~~ 2.624 &~~ 4$^+$  &~~ 2.609 &~~ 4$^+$  &~~ 2.550 &~~ 2$^+$  &~~ 2.444\\
   -   &~~  -    &~~ 4$^+$  &~~ 2.633 &~~ 1$^+$  &~~ 2.679 &~~ 1$^+$  &~~ 2.603 &~~ 3$^+$  &~~ 2.618\\
   -   &~~  -    &~~ 5$^+$  &~~ 2.652 &~~ 3$^+$  &~~ 2.702  &~~ 5$^+$  &~~ 2.613 &~~ 1$^+$  &~~ 2.638 \\
   -   &~~   -   &~~ 3$^+$  &~~ 2.672 &~~ 5$^+$  &~~ 2.780  &~~ 3$^+$  &~~ 2.629 &~~ 5$^+$  &~~ 2.699\\
   -   &~~   -   &~~ 6$^+$  &~~ 2.797 &~~ 6$^+$  &~~ 2.861&~~ 6$^+$  &~~ 2.809 &~~ 6$^+$  &~~ 2.963 \\
   -   &~~   -   &~~ 5$^+$  &~~ 3.038 &~~ 3$^+$  &~~ 3.074 &~~ 5$^+$  &~~ 2.979 &~~ 5$^+$  &~~ 3.052 \\
   -   &~~   -   &~~ 3$^+$  &~~ 3.061 &~~ 5$^+$  &~~ 3.083&~~ 3$^+$  &~~ 3.018 &~~ 3$^+$  &~~ 3.138 \\
   -   &~~   -   &~~ 1$^+$  &~~ 3.345 &~~ 8$^+$  &~~ 3.505&~~ 1$^+$  &~~ 3.292 &~~ 8$^+$  &~~ 3.464 \\
   -   &~~   -   &~~ 8$^+$  &~~ 3.623 &~~ 1$^+$  &~~ 3.543&~~ 8$^+$  &~~ 3.562 &~~ 1$^+$  &~~ 3.511\\
   -   &~~   -   &~~ 7$^+$  &~~ 3.946 &~~ 7$^+$  &~~ 3.751 &~~ 10$^+$  &~~ 3.912 &~~ 7$^+$  &~~ 3.702 \\
   -   &~~   -   &~~ 10$^+$  &~~ 3.995 &~~ 10$^+$  &~~ 3.826 &~~ 7$^+$  &~~ 3.917 &~~ 10$^+$  &~~ 3.878\\
\hline

\end{tabular}}
\end{center}
\end{table}

\begin{figure}
\begin{center}
\includegraphics[width=10.5cm]{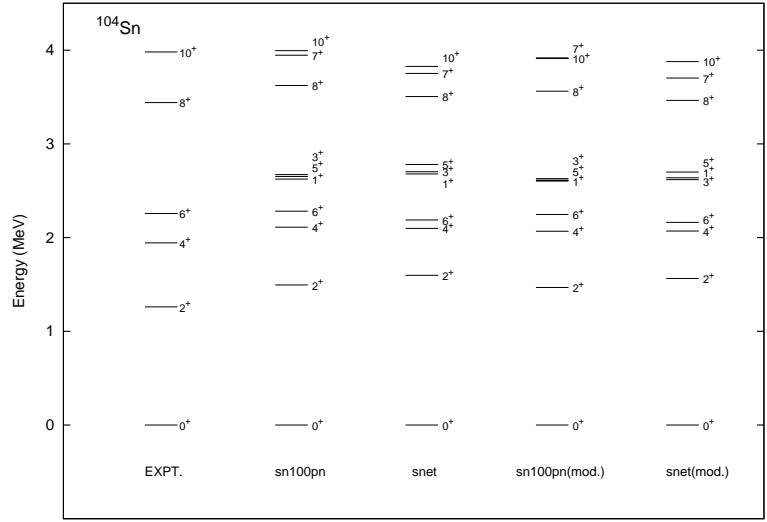}% Here is how to import EPS art
\caption{\label{104Sn}  Same as figure 1 for $^{104}$Sn.}
\end{center}
\end{figure}

The calculated and experimentally known energy levels for $^{104}$Sn are tabulated in Table 3 and are also plotted in Fig.\ \ref{104Sn}. Both the interactions predict very well the ordering of levels. In order to interpret the experimental results of $^{104}$Sn, G\'orska {\it et al.} \cite {Gor98} have also done similar calculations using snet interaction. Along with result of same interaction, we are also reporting results from sn100pn interaction for systematic study of this nucleus. The first excited 2$^+$ state is overestimated by 234, 337 keV from sn100pn and snet interaction from the experimental value, which is reduced by 207 and 303 keV, respectively when  single particle energies are adjusted in these interactions. The first 4$^+$ is predicted at 2111 and 2098 keV by sn100pn and snet interaction while corresponding experimental value is 1943 keV. The first 6$^+$ is predicted by both interactions within 70 keV difference from experimental observed data. The spacing between 4$^+$ and 6$^+$ gets compressed from snet interaction which remains almost unchanged with modified single particle energies too. The high-spin states 8$^+$ and 10$^+$ are calculated at 3623 and 3995 keV for sn100pn interaction whereas, from sn100pn (mod.) these states lie at 3562 and 3912 keV, which are closer to experimental data. With snet interaction, these states lie at 3505 and 3826 keV. The snet(mod.)
interaction predicts 8$^+$ and 10$^+$ at 3702, 3878, keV respectively.

\begin{table}
\begin{center}
\caption{
Experimental \protect\cite{Juu97} and theoretical low-lying states (up to 4.5 MeV) of $^{106}$Sn. Energies are in MeV.}
\label{106Sn2}
\resizebox{12.7cm}{!}{
\begin{tabular}{rcrcrcrcrcrc}
\hline

J$^\pi$ & ~~Exp. &~~ J$^\pi$ &~~ sn100pn &~~  J$^\pi$ &~~ snet~~ &~~ J$^\pi$ &~~ sn100pn(mod.) &~~  J$^\pi$ &~~ snet(mod.)~~\\	
\hline
(0$^+$) &~~ 0.000 &~~ 0$^+$  &~~ 0.000 &~~ 0$^+$ &~~ 0.000  &~~ 0$^+$  &~~ 0.000 &~~ 0$^+$ &~~ 0.000\\
(2$^+$) &~~ 1.208 &~~ 2$^+$  &~~ 1.413 &~~ 2$^+$ &~~ 1.558  &~~ 2$^+$  &~~ 1.399 &~~ 2$^+$ &~~ 1.569\\
(4$^+$) &~~ 2.202 &~~ 4$^+$  &~~ 2.194 &~~ 4$^+$ &~~ 2.167 &~~ 4$^+$  &~~ 2.170 &~~ 0$^+$ &~~ 2.138 \\
(6$^+$) &~~ 2.324 &~~ 6$^+$  &~~ 2.424 &~~ 6$^+$ &~~ 2.306 &~~ 0$^+$  &~~ 2.368 &~~ 4$^+$ &~~ 2.140\\
(8$^+$) &~~ 3.480 &~~ 0$^+$  &~~ 2.430 &~~ 0$^+$ &~~ 2.361 &~~ 6$^+$  &~~ 2.383 &~~ 6$^+$ &~~ 2.281 \\
(10$^+$) &~~ 4.133 &~~ 2$^+$  &~~ 2.592 &~~ 4$^+$ &~~ 2.615 &~~ 2$^+$  &~~ 2.519 &~~ 4$^+$ &~~ 2.514 \\
   -   &~~  -     &~~ 4$^+$  &~~ 2.676 &~~ 2$^+$  &~~ 2.625  &~~ 4$^+$  &~~ 2.592 &~~ 2$^+$  &~~ 2.573\\
   -   &~~  -    &~~ 1$^+$  &~~ 2.723 &~~ 1$^+$  &~~ 2.771 &~~ 1$^+$  &~~ 2.672 &~~ 5$^+$  &~~ 2.704 \\
   -   &~~  -    &~~ 5$^+$  &~~ 2.762 &~~ 3$^+$  &~~ 2.799 &~~ 5$^+$  &~~ 2.703 &~~ 3$^+$  &~~ 2.709 \\
   -   &~~   -   &~~ 3$^+$  &~~ 2.766 &~~ 5$^+$  &~~ 2.822 &~~ 3$^+$  &~~ 2.711 &~~ 1$^+$  &~~ 2.741\\
   -   &~~   -   &~~ 6$^+$  &~~ 2.985 &~~ 6$^+$  &~~ 2.932 &~~ 6$^+$  &~~ 2.967 &~~ 6$^+$  &~~ 2.897\\
   -   &~~   -   &~~ 5$^+$  &~~ 3.080 &~~ 5$^+$  &~~ 3.169 &~~ 5$^+$  &~~ 3.036 &~~ 5$^+$  &~~ 3.214\\
   -   &~~   -   &~~ 1$^+$  &~~ 3.481 &~~ 3$^+$  &~~ 3.293 &~~ 3$^+$  &~~ 3.111 &~~ 3$^+$  &~~ 3.298\\
   -   &~~   -   &~~ 8$^+$  &~~ 3.652 &~~ 8$^+$  &~~ 3.660 &~~ 1$^+$  &~~ 3.435 &~~ 1$^+$  &~~ 3.619\\
   -   &~~   -   &~~ 7$^+$  &~~ 4.020 &~~ 1$^+$  &~~ 3.692 &~~ 8$^+$  &~~ 3.586 &~~ 8$^+$  &~~ 3.660\\
   -   &~~   -   &~~ 7$^+$  &~~ 4.119 &~~7$^+$  &~~ 3.797 &~~ 7$^+$  &~~ 3.936 &~~7$^+$  &~~ 3.756 \\
   -   &~~   -   &~~ 10$^+$  &~~ 4.362 &~~ 10$^+$ &~~ 4.125 &~~ 10$^+$  &~~ 4.300 &~~10$^+$  &~~ 4.086\\
\hline

\end{tabular}}
\end{center}
\end{table}

\begin{figure}
\begin{center}
\includegraphics[width=10.5cm]{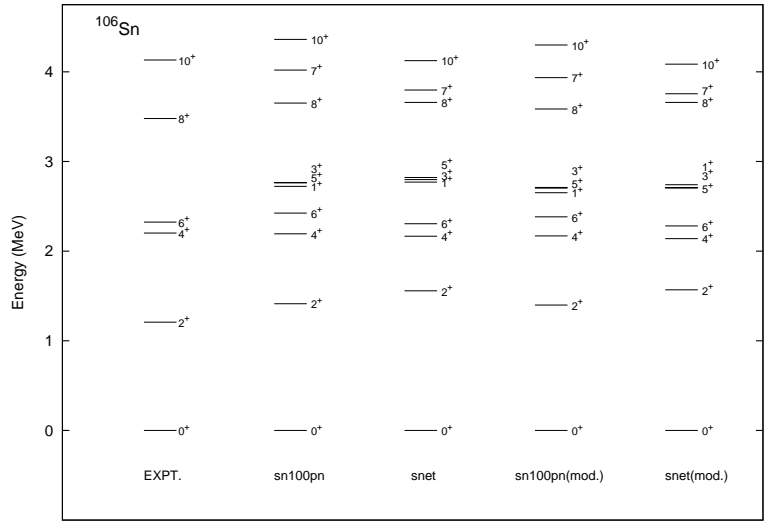}% Here is how to import EPS art
\caption{\label{106Sn}  Same as figure 1 for $^{106}$Sn.}
\end{center}
\end{figure}

For $^{106}$Sn, the calculated and experimental states are listed in Table 4  and corresponding energy spectra are shown in Fig.\ \ref{106Sn}. More recently shell model calculation has been done by Dikmen \cite {Dik09} for $^{106}$Sn using CD Bonn potential. The calculation reported in this paper using two different types of interactions have some what over estimated the experimentally observed ones, while our calculation using sn100pn and snet interactions are more close to the experimentally observed data. The position of 2$^+$ level at 1.413 MeV from sn100pn interaction agrees with experimentally observed one at 1.208 MeV but it is over estimated by 350 keV in snet interaction. Again the 4$^+$, 6$^+$, 8$^+$ and 10$^+$ states are in good agreement with experimental data from both the interaction, even in much better agreement with snet interaction. However, changing the single particle energies in both the interaction does not show significant improvement in the energy spectrum for these levels.

\begin{table}
\begin{center}
\caption{
Experimental \protect\cite {Wad96} and theoretical low-lying states (up to 4.5 MeV) of $^{108}$Sn. Energies are in MeV.}
\label{108Sn}
\resizebox{12.7cm}{!}{
\begin{tabular}{rcrcrcrcrcrc}
\hline

J$^\pi$ & ~~Exp. &~~ J$^\pi$ &~~ sn100pn &~~  J$^\pi$ &~~ snet~~ &~~ J$^\pi$ &~~ sn100pn(mod.) &~~  J$^\pi$ &~~ snet(mod.)~~\\			
\hline
(0$^+$) &~~ 0.000 &~~ 0$^+$  &~~ 0.000 &~~ 0$^+$ &~~ 0.000 &~~ 0$^+$  &~~ 0.000 &~~ 0$^+$ &~~ 0.000 \\
(2$^+$) &~~ 1.207 &~~ 2$^+$  &~~ 1.186 &~~ 2$^+$ &~~ 1.498   &~~ 2$^+$  &~~ 1.201 &~~ 2$^+$ &~~ 1.540  \\
(4$^+$) &~~ 2.111 &~~ 4$^+$  &~~ 2.142 &~~ 4$^+$ &~~ 2.177 &~~ 4$^+$  &~~ 2.146&~~ 4$^+$ &~~ 2.146  \\
(6$^+$) &~~ 2.365 &~~ 2$^+$  &~~ 2.264 &~~ 6$^+$ &~~ 2.315 &~~ 0$^+$  &~~ 2.247 &~~ 0$^+$ &~~ 2.163\\
(8$^+$) &~~ 3.561 &~~ 6$^+$  &~~ 2.297 &~~ 0$^+$ &~~ 2.326 &~~ 2$^+$  &~~ 2.248 &~~ 6$^+$ &~~ 2.299 \\
(10$^+$) &~~ 4.256 &~~ 0$^+$  &~~ 2.315 &~~ 2$^+$ &~~ 2.606 &~~ 6$^+$  &~~ 2.283 &~~ 2$^+$ &~~ 2.526 \\
   -   &~~  -     &~~ 4$^+$  &~~ 2.338 &~~ 4$^+$  &~~ 2.607 &~~ 4$^+$  &~~ 2.316 &~~ 4$^+$  &~~2.589 \\
   -   &~~  -    &~~ 1$^+$  &~~ 2.575 &~~ 5$^+$  &~~ 2.767 &~~ 5$^+$  &~~ 2.521 &~~ 5$^+$  &~~  2.674\\
   -   &~~  -    &~~ 5$^+$  &~~ 2.588 &~~ 1$^+$  &~~ 2.771 &~~ 3$^+$  &~~ 2.560 &~~ 3$^+$  &~~ 2.726\\
   -   &~~   -   &~~ 3$^+$  &~~ 2.607 &~~ 3$^+$  &~~ 2.797 &~~ 1$^+$  &~~ 2.582 &~~ 1$^+$  &~~ 2.756 \\
   -   &~~   -   &~~ 6$^+$  &~~ 2.985 &~~ 6$^+$  &~~ 2.859 &~~ 5$^+$  &~~ 2.794 &~~ 6$^+$  &~~ 2.762\\
   -   &~~   -   &~~ 5$^+$  &~~ 2.759 &~~ 5$^+$  &~~ 3.115 &~~ 6$^+$  &~~ 2.816 &~~ 1$^+$  &~~ 3.165\\
   -   &~~   -   &~~ 6$^+$  &~~ 2.840 &~~ 3$^+$  &~~ 3.338 &~~ 3$^+$  &~~ 2.909 &~~ 5$^+$  &~~ 3.208 \\
   -   &~~   -   &~~ 3$^+$  &~~ 2.886 &~~ 7$^+$  &~~ 3.666&~~ 1$^+$  &~~ 2.951 &~~3$^+$  &~~ 3.312 \\
   -   &~~   -   &~~ 1$^+$  &~~ 3.006 &~~ 8$^+$  &~~ 3.693 &~~ 8$^+$  &~~ 3.500 &~~8$^+$  &~~ 3.672\\
   -   &~~   -   &~~ 8$^+$  &~~ 3.502 &~~ 8$^+$  &~~ 4.054 &~~ 7$^+$  &~~ 3.608 &~~ 7$^+$  &~~ 3.689 \\
   -   &~~   -   &~~ 7$^+$  &~~ 3.612 &~~ 7$^+$ &~~ 4.137 &~~ 8$^+$  &~~ 3.889 &~~ 8$^+$ &~~ 4.023 \\
   -   &~~   -   &~~ 10$^+$  &~~ 4.051 &~~ 10$^+$ &~~ 4.235&~~ 10$^+$  &~~ 4.199 &~~ 10$^+$ &~~ 4.211 \\
\hline

\end{tabular}}
\end{center}
\end{table}

\begin{figure}
\begin{center}
\includegraphics[width=10.5cm]{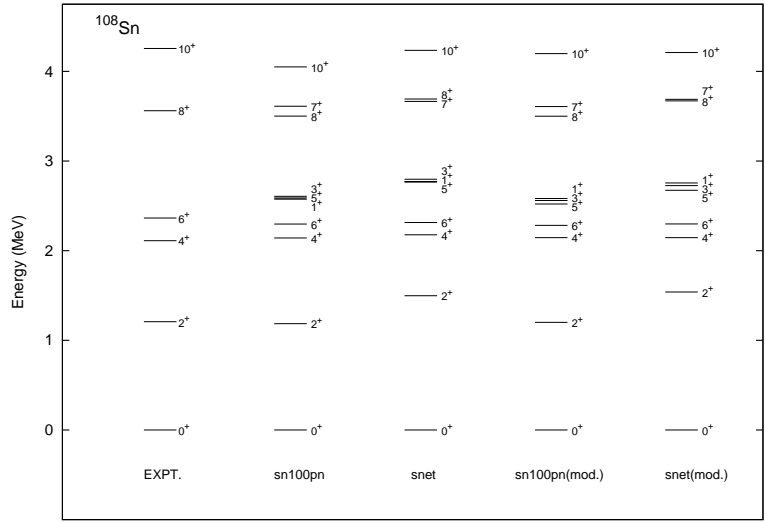}% Here is how to import EPS art
\caption{\label{108Sn}  Same as figure 1 for $^{108}$Sn.}
\end{center}
\end{figure}

Table 5, contains experimentally observed and theoretically calculated states up to 4.5 MeV for $^{108}$Sn. These states are also shown in Fig.\ \ref{108Sn}. In both set of shell model calculations for $^{108}$Sn, the calculations have been performed  by imposing restriction of maximum two particles in h$_{11/2}$ orbital to tackle dimension of matrices.
For the sn100pn interaction, the calculated 2$^+$, 4$^+$, 6$^+$, 8$^+$ and 10$^+$ states are  predicted at 1186, 2142, 2297, 3502 and 4051 keV corresponding to experimental value of 1207, 2111, 2365, 3561 and 4256 keV. It has been observed that experimental levels are in much better agreement from sn100pn as well as snet interactions as compared to the previous work done by Dikmen\cite {Dik09}. 
In general, the calculations performed with new single particle energies \cite{Darby10}, significantly improve the predictive power of calculations as compared to the calculations performed with previously taken single particle energies.

\subsection{Excitation energies of odd Sn isotopes}

The calculated energy spectra for odd $^{103-107}$Sn isotopes are shown in Figs.5-6. In the present calculations the energy spectra are plotted up to 2.5 MeV excitation energy whereas previous calculations were reported up to 2 MeV. Here, we have extended the calculations with better agreement from experimental data for odd A Sn isotopes. Some characteristic features have been reported ahead.

\begin{table}
\begin{center}
\caption{
Experimental \protect\cite {Fah01} and theoretical low-lying states (up to 2.5 MeV) of $^{103}$Sn. Energies are in MeV.}
\label{t_mn62}
\resizebox{12.7cm}{!}{
\begin{tabular}{rcrcrcrcrcrc}
\hline

J$^\pi$ & ~~Exp. &~~ J$^\pi$ &~~ sn100pn &~~  J$^\pi$ &~~ snet~~ &~~ J$^\pi$ &~~ sn100pn(mod.) &~~  J$^\pi$ &~~ snet(mod.)~~\\			
\hline
(5/2$^+$) &~~ 0.000 &~~ 5/2$^+$  &~~ 0.000 &~~ 5/2$^+$ &~~ 0.000 &~~ 5/2$^+$  &~~ 0.000 &~~ 5/2$^+$ &~~ 0.000  \\
(7/2$^+$) &~~ 0.168 &~~ 7/2$^+$  &~~ 0.026 &~~ 7/2$^+$ &~~ 0.077 &~~ 7/2$^+$  &~~ 0.107 &~~ 7/2$^+$ &~~ 0.254 \\
(11/2$^+$) &~~ 1.486 &~~ 5/2$^+$  &~~ 1.124 &~~ 3/2$^+$ &~~ 1.008 &~~ 1/2$^+$  &~~ 1.267 &~~ 3/2$^+$ &~~ 0.920 \\
(13/2$^+$) &~~ 1.785 &~~ 1/2$^+$  &~~ 1.186 &~~ 5/2$^+$ &~~ 1.241 &~~ 5/2$^+$  &~~ 1.268 &~~ 5/2$^+$ &~~ 1.351  \\
    - &~~ - &~~ 7/2$^+$  &~~ 1.282 &~~ 3/2$^+$ &~~ 1.284 &~~ 7/2$^+$  &~~ 1.304 &~~ 9/2$^+$ &~~ 1.374 \\
  -  &~~  - &~~ 3/2$^+$  &~~ 1.307 &~~ 9/2$^+$ &~~ 1.399 &~~ 3/2$^+$  &~~ 1.325 &~~ 3/2$^+$ &~~ 1.420 \\
  -  &~~  - &~~ 3/2$^+$  &~~ 1.355 &~~ 1/2$^+$ &~~ 1.512 &~~ 9/2$^+$  &~~ 1.430 &~~ 1/2$^+$ &~~ 1.550  \\
  -  &~~ - &~~ 9/2$^+$  &~~ 1.395 &~~ 7/2$^+$ &~~ 1.521 &~~ 3/2$^+$  &~~ 1.435 &~~ 9/2$^+$ &~~ 1.587 \\
  -  &~~  &~~ 11/2$^+$  &~~ 1.635 &~~ 11/2$^+$ &~~ 1.601 &~~ 11/2$^+$  &~~ 1.647 &~~ 11/2$^+$ &~~ 1.668 \\
   -   &~~  -     &~~ 9/2$^+$  &~~ 1.717 &~~ 9/2$^+$  &~~ 1.612 &~~ 9/2$^+$  &~~ 1.663 &~~ 1/2$^+$  &~~ 1.865 \\
   -   &~~  -    &~~ 1/2$^+$  &~~ 1.749 &~~ 1/2$^+$  &~~ 1.660  &~~ 1/2$^+$  &~~ 1.784 &~~ 15/2$^+$  &~~ 1.885 \\
   -   &~~  -    &~~ 11/2$^+$  &~~ 1.763 &~~ 11/2$^+$  &~~ 1.751 &~~ 11/2$^+$  &~~ 1.799 &~~ 13/2$^+$  &~~ 1.904 \\
   -   &~~   -   &~~ 13/2$^+$  &~~ 1.798 &~~ 17/2$^+$  &~~ 1.925 &~~ 15/2$^+$  &~~ 1.839 &~~ 11/2$^+$  &~~ 1.925\\
   -   &~~   -   &~~ 17/2$^+$  &~~ 1.876 &~~ 13/2$^+$  &~~ 1.931 &~~ 13/2$^+$  &~~ 1.840 &~~ 13/2$^+$  &~~ 1.982 \\
   -   &~~   -   &~~ 15/2$^+$  &~~ 1.924 &~~ 15/2$^+$  &~~ 1.938 &~~ 17/2$^+$  &~~ 1.918 &~~ 17/2$^+$  &~~ 2.085\\
   -   &~~   -   &~~ 15/2$^+$  &~~ 2.066 &~~ 15/2$^+$  &~~ 2.010 &~~ 13/2$^+$  &~~ 2.079 &~~ 15/2$^+$  &~~ 2.252\\
\hline

\end{tabular}}
\end{center}
\end{table}

\begin{figure}
\begin{center}
\includegraphics[width=10.5cm]{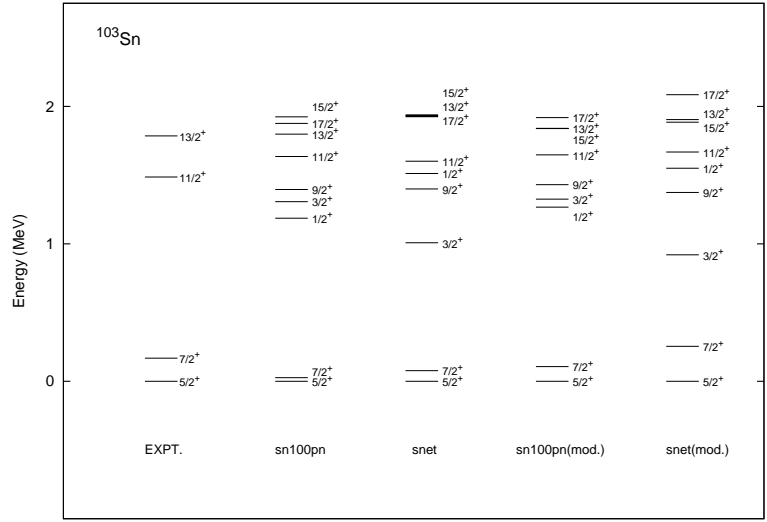}% Here is how to import EPS art
\caption{\label{103Sn}Experimental \protect\cite {Fah01} and calculated energy levels of $^{103}$Sn with sn100pn and snet interactions.}
\end{center}
\end{figure}

In Table 6, we have listed all experimentally \cite {Fah01} known and calculated low-lying states for $^{103}$Sn. Experimentally, only three excited states 7/2$^+$, 11/2$^+$ and 13/2$^+$ are known above the ground state 5/2$^+$. From Fig.\ \ref{103Sn} we can see that in both sets of calculations, the splitting of two low-lying states 5/2$^+$ and 7/2$^+$, which are essentially single particle in nature, is under estimated by $\sim$140 and $\sim$90 keV from sn100pn and snet interactions whereas, from recently observed single particle energies \cite{Darby10}, this improves with the difference of $\sim$ 60 keV and $\sim$ 85 keV, respectively from both interactions. Above 7/2$^+$, experimentally their is large pronounced gap of about 1.3 MeV from the first 11/2$^+$ state. This is well reproduced by both the interactions. Our calculations predict many more levels as compared to the levels observed experimentally.
\begin{figure}
\begin{center}
\includegraphics[width=10.5cm]{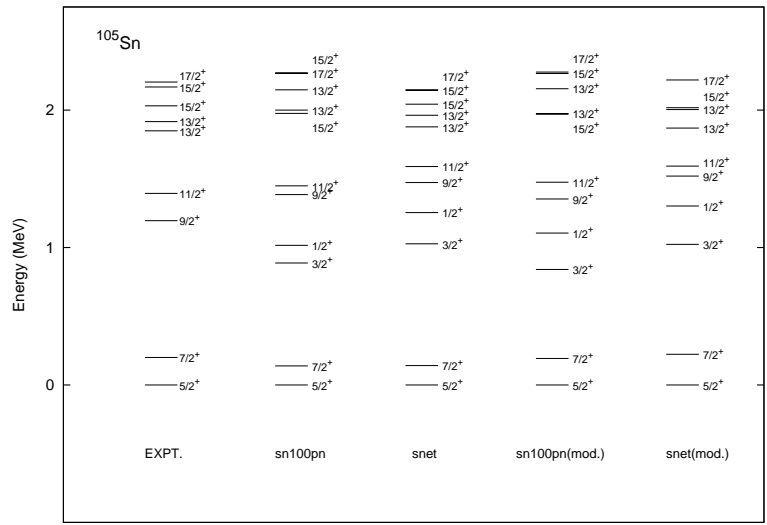}% Here is how to import EPS art
\caption{\label{105Sn}  Same as figure 1 for $^{105}$Sn.}
\end{center}
\end{figure}
\begin{table}
\begin{center}
\caption{
Experimental \protect\cite {Gad97} and theoretical low-lying states (up to 2.5 MeV) of $^{105}$Sn. Energies are in MeV.}
\label{t_mn62}
\resizebox{12.7cm}{!}{
\begin{tabular}{rcrcrcrcrcrc}
\hline

J$^\pi$ & ~~Exp. &~~ J$^\pi$ &~~ sn100pn &~~  J$^\pi$ &~~ snet~~ &~~ J$^\pi$ &~~ sn100pn(mod.) &~~  J$^\pi$ &~~ snet(mod.)~~\\			
\hline
(5/2$^+$) &~~ 0.000 &~~ 5/2$^+$  &~~ 0.000 &~~ 5/2$^+$ &~~ 0.000 &~~ 5/2$^+$  &~~ 0.000 &~~ 5/2$^+$ &~~ 0.000  \\
(7/2$^+$) &~~ 0.200 &~~ 7/2$^+$  &~~ 0.139 &~~ 7/2$^+$ &~~ 0.141  &~~ 7/2$^+$  &~~0.193 &~~ 7/2$^+$ &~~ 0.223 \\
(9/2$^+$) &~~ 1.195 &~~ 3/2$^+$  &~~ 0.887 &~~ 3/2$^+$ &~~ 1.027  &~~ 3/2$^+$  &~~ 0.840 &~~ 3/2$^+$ &~~ 1.023\\
(11/2$^+$) &~~ 1.394 &~~ 5/2$^+$  &~~ 0.942 &~~ 3/2$^+$ &~~ 1.144 &~~ 5/2$^+$  &~~ 0.998 &~~ 3/2$^+$ &~~ 1.172 \\
(13/2$^+$) &~~ 1.849 &~~ 1/2$^+$  &~~ 1.016 &~~ 5/2$^+$ &~~ 1.199 &~~ 1/2$^+$  &~~ 1.105 &~~ 5/2$^+$ &~~ 1.253 \\
(13/2$^+$) &~~ 1.916 &~~ 3/2$^+$  &~~ 1.177 &~~ 1/2$^+$ &~~ 1.254 &~~ 3/2$^+$  &~~ 1.210 &~~ 1/2$^+$ &~~ 1.302 \\
(15/2$^+$) &~~ 2.031 &~~ 9/2$^+$  &~~ 1.385 &~~ 1/2$^+$ &~~ 1.450 &~~ 9/2$^+$  &~~ 1.353 &~~ 7/2$^+$ &~~ 1.506 \\
(15/2$^+$) &~~ 2.168 &~~ 11/2$^+$  &~~ 1.449 &~~ 9/2$^+$ &~~ 1.473 &~~ 11/2$^+$  &~~ 1.475 &~~ 9/2$^+$ &~~ 1.520 \\
(17/2$^+$) &~~ 2.204 &~~ 1/2$^+$  &~~ 1.501 &~~ 7/2$^+$ &~~ 1.500  &~~ 1/2$^+$  &~~ 1.536 &~~ 11/2$^+$ &~~ 1.592 \\
   -   &~~  -     &~~ 7/2$^+$  &~~ 1.553 &~~ 9/2$^+$  &~~ 1.579 &~~ 7/2$^+$  &~~ 1.558 &~~ 9/2$^+$  &~~ 1.598\\
   -   &~~  -    &~~ 9/2$^+$  &~~ 1.692 &~~11/2$^+$  &~~ 1.589 &~~ 9/2$^+$  &~~ 1.708 &~~1/2$^+$  &~~ 1.628 \\
   -   &~~  -    &~~ 15/2$^+$  &~~ 1.976 &~~ 13/2$^+$  &~~ 1.878 &~~ 15/2$^+$  &~~ 1.970 &~~ 13/2$^+$  &~~ 1.869 \\
   -   &~~   -   &~~ 13/2$^+$  &~~ 2.000 &~~ 13/2$^+$  &~~ 1.962 &~~ 13/2$^+$  &~~ 1.974 &~~ 13/2$^+$  &~~ 2.004\\
   -   &~~   -   &~~ 11/2$^+$  &~~ 2.066 &~~ 11/2$^+$  &~~ 2.017  &~~ 11/2$^+$  &~~ 2.070 &~~ 15/2$^+$  &~~ 2.017 \\
   -   &~~   -   &~~ 13/2$^+$  &~~ 2.147 &~~ 15/2$^+$  &~~ 2.043 &~~ 13/2$^+$  &~~ 2.155 &~~ 11/2$^+$  &~~ 2.050\\
   -   &~~   -   &~~ 17/2$^+$  &~~ 2.267 &~~ 15/2$^+$  &~~ 2.143 &~~ 15/2$^+$  &~~ 2.262&~~ 15/2$^+$  &~~ 2.219\\
   -   &~~   -   &~~ 15/2$^+$  & 2.271 &~~ 17/2$^+$  &~~ 2.147 &~~ 17/2$^+$   & 2.276 &~~ 17/2$^+$  &~~ 2.219 \\
\hline

\end{tabular}}
\end{center}
\end{table}

 Fig.\ \ref{105Sn}, compares the experimental\cite {Gad97} and theoretical low lying states of $^{105}$Sn and corresponding excitation energies are  tabulated in Table 7.  Experimentally, the first 7/2$^+$ state lies at 200 keV above the 5/2$^+$ ground state, which is reproduced very well with the modified single particle energies in both the interactions. The levels observed above the 7/2$^+$ are reproduced by both interactions within  reasonable accuracy. In comparison to the previous work by Schubert {\it et al.} \cite {Sch95}, our calculations predict much better spectra.

\begin{figure}
\begin{center}
\includegraphics[width=10.5cm]{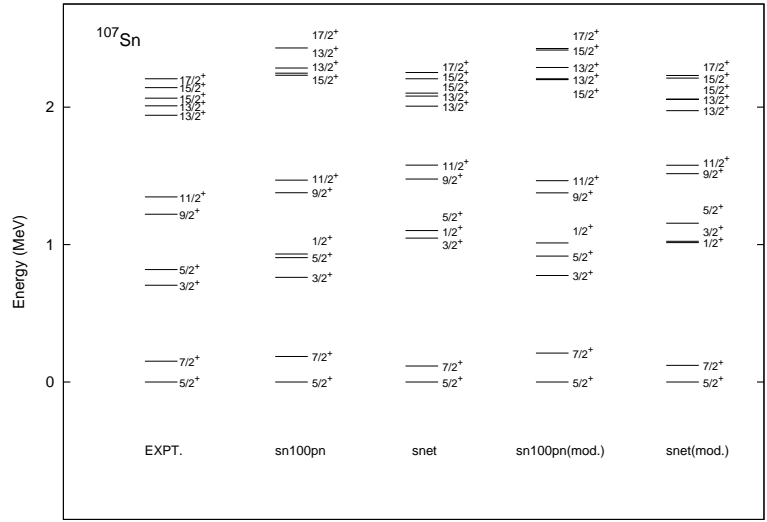}
%\resizebox{0.5\textwidth}{!}{
%  \includegraphics{sn107.eps}
%}
%\includegraphics[width=8cm]{sn107.eps}% Here is how to import EPS art
\caption{\label{107Sn} Same as figure 1 for $^{107}$Sn.}
\end{center}
\end{figure}
\begin{table}
\begin{center}
\caption{
Experimental \protect\cite{Ish93,Res02} and theoretical low-lying states (up to 2.5 MeV) of $^{107}$Sn. Energies are in MeV.}
\label{t_mn62}
\resizebox{12.7cm}{!}{
\begin{tabular}{rcrcrcrcrcrc}
\hline

J$^\pi$ & ~~Exp. &~~ J$^\pi$ &~~ sn100pn &~~  J$^\pi$ &~~ snet~~ &~~ J$^\pi$ &~~ sn100pn(mod.) &~~  J$^\pi$ &~~ snet(mod.)~~\\			
\hline
(5/2$^+$) &~~ 0.000 &~~ 5/2$^+$  &~~ 0.000 &~~ 5/2$^+$ &~~ 0.000 &~~ 5/2$^+$  &~~ 0.000 &~~ 5/2$^+$ &~~ 0.000 \\
(7/2$^+$) &~~ 0.151 &~~ 7/2$^+$  &~~ 0.186 &~~ 7/2$^+$ &~~ 0.116 &~~ 7/2$^+$  &~~ 0.210 &~~ 7/2$^+$ &~~ 0.121 \\
(3/2$^+$) &~~ 0.704 &~~ 3/2$^+$  &~~ 0.761 &~~ 3/2$^+$ &~~ 1.002 &~~ 3/2$^+$  &~~ 0.775&~~ 1/2$^+$ &~~ 1.014 \\
(5/2$^+$) &~~ 0.818 &~~ 5/2$^+$  &~~ 0.905 &~~ 1/2$^+$ &~~ 1.047 &~~ 5/2$^+$  &~~ 0.916&~~ 3/2$^+$ &~~ 1.024 \\
(3/2$^+$) &~~ 0.970 &~~ 1/2$^+$  &~~ 0.931 &~~ 5/2$^+$ &~~ 1.102 &~~ 1/2$^+$  &~~ 1.012 &~~ 5/2$^+$ &~~ 1.155 \\
(9/2$^+$) &~~ 1.221 &~~ 3/2$^+$  &~~ 1.133 &~~ 3/2$^+$ &~~ 1.148 &~~ 3/2$^+$  &~~ 1.166 &~~ 3/2$^+$ &~~ 1.202 \\
(3/2$^+$) &~~ 1.280 &~~ 3/2$^+$  &~~ 1.284 &~~ 7/2$^+$ &~~ 1.453 &~~ 9/2$^+$  &~~ 1.376 &~~ 7/2$^+$ &~~ 1.466 \\
(11/2$^+$) &~~ 1.347 &~~ 9/2$^+$  &~~ 1.377 &~~ 9/2$^+$ &~~ 1.477 &~~ 7/2$^+$  &~~ 1.450 &~~ 9/2$^+$ &~~ 1.516 \\
(3/2$^+$) &~~ 1.454 &~~ 7/2$^+$  &~~ 1.445 &~~ 11/2$^+$ &~~ 1.578 &~~ 11/2$^+$  &~~ 1.465 &~~ 1/2$^+$ &~~ 1.571 \\
(13/2$^+$) &~~ 1.941 &~~ 11/2$^+$  &~~ 1.469 &~~ 9/2$^+$ &~~ 1.579 &~~ 1/2$^+$  &~~ 1.476 &~~ 11/2$^+$ &~~ 1.577 \\
(13/2$^+$) &~~ 2.009 &~~ 9/2$^+$  &~~ 1.638 &~~ 3/2$^+$ &~~ 1.709  &~~ 9/2$^+$  &~~ 1.647 &~~ 9/2$^+$ &~~ 1.630\\
(15/2$^+$) &~~ 2.065 &~~ 3/2$^+$  &~~ 1.780 &~~ 3/2$^+$ &~~ 1.740 &~~ 11/2$^+$  &~~ 2.172 &~~ 11/2$^+$ &~~ 1.950 \\
(15/2$^+$) &~~ 2.142 &~~ 11/2$^+$  &~~ 2.189 &~~ 13/2$^+$ &~~ 2.007 &~~ 15/2$^+$  &~~ 2.200 &~~ 13/2$^+$ &~~ 1.974\\
(17/2$^+$) &~~ 2.206 &~~ 15/2$^+$  &~~ 2.232 &~~ 13/2$^+$ &~~ 2.080 &~~ 13/2$^+$  &~~ 2.206 &~~ 13/2$^+$ &~~ 2.056 \\
   -   &~~  -     &~~ 13/2$^+$  &~~ 2.247 &~~ 15/2$^+$  &~~ 2.102 &~~ 13/2$^+$  &~~ 2.288 &~~ 15/2$^+$  &~~ 2.058\\
   -   &~~  -    &~~ 13/2$^+$  &~~ 2.284 &~~15/2$^+$  &~~ 2.226  &~~ 15/2$^+$  &~~ 2.415 &~~15/2$^+$  &~~ 2.211 \\
   -   &~~  -    &~~ 17/2$^+$  &~~ 2.430 &~~ 17/2$^+$  &~~ 2.252 &~~ 17/2$^+$  &~~ 2.426 &~~ 17/2$^+$  &~~ 2.230 \\
   -   &~~   -   &~~ 15/2$^+$  &~~ 2.438 &~~ -  &~~ -&~~ 17/2$^+$  &~~ 3.492 &~~ 17/2$^+$  &~~ 3.145 \\
\hline

\end{tabular}}
\end{center}
\end{table}

In Table 8, we have listed experimental and theoretical low lying states of $^{107}$Sn and corresponding levels are drawn in Fig.\ \ref{107Sn}.
We have extended the calculations performed by Dikmen \cite {Dik09} to higher spin states. Dikmen has performed the shell model calculation for low lying levels using CD Bonn potential and Nijmenjal1 interactions. In our calculations, using sn100pn interaction we have better agreement for low-lying states, whereas for levels above 11/2$^+$, ordering is well reproduced but values are slightly over predicted. However, calculations performed using snet interaction are able to explain low-lying states as well as higher-lying levels with better accuracy. Hence it is required to retune the TBME or readjust the single particle energies in sn100pn interaction.

\section{Electromagnetic transition strengths}
The reduced transition probabilities $ B(E2;0^{+}_{g.s}\rightarrow 2^{+}_{1})$, obtained from calculation with the experimental value
\cite{Ban05,Vam07} are shown in Table 9. The neutron effective charge used in the present calculation is 1.0e, same as used in Refs. \cite{Ban05,Vam07}. The results obtained from both the interactions have almost similar pattern and increase in B(E2) values have been observed with increasing neutron number, which is a signature for enhancement in collectivity.

\begin{table}
\begin{center}
\caption{Comparison of the B(E2; $0^{+}_{g.s}$$\rightarrow$$2^{+}_{1}$) values, obtained from the shell model calculations with the experimental values in e$^2$b$^2$ for even Sn isotopes. The experimental values have been taken from \protect\cite{Ban05,Vam07}.}
\label{be2}
%\resizebox{7.5cm}{3.5cm}
\begin{tabular}{rcrcrcrc}

\hline\\
 Isotope &&$\emph{B}$($\emph{E2}$, $0^{+}_{g.s}$$\rightarrow$$2^{+}_{1}$)[e$^2$b$^2$]&&\\
 & ~~ Exp. &~~ sn100pn &~~ snet \\
\hline
 $^{102}$Sn  &~~ -    &~~ 0.039 &~~ 0.041   \\
 $^{104}$Sn  &~~  -   &~~ 0.090 &~~ 0.083 \\
 $^{106}$Sn  &~~  0.240   &~~ 0.138 &~~ 0.116   \\
 $^{108}$Sn  &~~  0.230  &~~ 0.176 &~~ -  \\
 \hline
\end{tabular}
\end{center}
\end{table}
\section{\label{Conclusion} Conclusion}
The shell model calculations for Sn isotopes with $A$ =102-108, have been performed in $sdg_{7/2}h_{11/2}$ valence space with two different types of interactions. Theoretical calculations show better agreement for even Sn isotopes with experimental data as compared to odd Sn isotopes. In general, the present calculations are in better agreement with experimental data with sn100pn interactions as compared to the snet interaction.
However, the results of sn100pn interaction with modified single particle energies improves the calculations significantly.
It is observed that overall agreement between theory and experimental data are quite satisfactory. Further, more experimental data is needed for lighter Sn isotopes for making more complete assessment of the calculations.

\vspace{0.5cm}

\section*{Acknowledgements}
Discussions with Dr. Sandeep Ghugre, UGC-DAE-CSR, Kolkata, India, is gratefully acknowledged.
P.C.S. acknowledges support from Conacyt, M´exico, and by
DGAPA, UNAM project IN103212.


\begin{thebibliography}{99}
\bibitem{Ots10} T. Otsuka \emph{et al.,}  Phys. Rev. Lett. {\bf 104} (2010) 012501; A. Gade and T. Glasmacher,
Prog. Part. Nucl. Phys.  {\bf 60} (2008) 161; O. Sorlin and M.-G. Porquet,
Prog. Part. Nucl. Phys.  {\bf 61} (2008) 602.
\bibitem{Fed79} P. Federman and S. Pittel, Phys. Rev. {\bf C20} (1979) 0820.
\bibitem{Wad96} R. Wadsworth \emph{et al.,} Phys. Rev.{\bf C53} (1996) 2763.
\bibitem{Gad97} A. Gadea \emph{et al.,} Phys. Rev.{\bf C55} (1997) 1(R).
\bibitem{Fae02} T. Faestermann \emph{et al.,} Eur. Phys. J. {\bf A15} (2002) 185.
\bibitem{Fah01} C. Fahlander\emph{et al.,} Phys. Rev.{\bf C63} (1996) 021307(R).
\bibitem{Gor98} M. G\'orska \emph{et al.,} Phys. Rev.{\bf C58} (1998) 108.
\bibitem{Juu97} S. Juutinen \emph{et al.,} Nucl. Phys. {\bf A617} (1997) 74.
\bibitem{Ban05} A. Banu \emph{et al.,} Phys. Rev. {\bf C72} (2005) 061305(R).
\bibitem{Vam07} C. Vaman \emph{et al.,}  Phys. Rev. Lett. {\bf 99} (2007) 162501.
\bibitem{Res02} J. J. Ressler \emph{et al.,} Phys. Rev.{\bf C65} (2002) 044330.
\bibitem{Eng95} T. Engeland \emph{et al.,}  Phys. Scri.{\bf T56} 58 (1995).
\bibitem{Cov97} A. Covello \emph{et al.,} Prog Part. Nucl. Phys, Vol  {\bf 38} (1997) 162.
\bibitem{Sch95} R. Schubart \emph{et al.,} Z. Phys.  {\bf A352} (1995) 373.
\bibitem{Dik09} E. Dikmen  Commun. Theor. Phys. {\bf 51} (2009) 899.
\bibitem{Bro05} B. A Brown \emph{et al.,} Phys. Rev.  {\bf C71} (2005) 044317.
\bibitem{Mac96} R. Machleidt, F. Sammarruca, and Y.Song, Phys. Rev.  {\bf C53} (1996) R1483.
\bibitem{Darby10} I.G. Darby \emph{et al.,}  Phys. Rev. Lett. {\bf 105} (2010) 162502.
\bibitem{Hos85} A. Hosaka \emph{et al.,}  Nucl. Phys. {\bf A444} (1985) 76.
\bibitem{Bro94} B. A Brown \emph{et al.,} Phys. Rev.  {\bf C50} (1994) 2270(R).
\bibitem{Bro07} B. A Brown and W. D. M. Rae, Nushell@MSU, MSU-NSCL report (2007).
\bibitem{Lip98} M. Lipoglavsek \emph{et al.,} Phys. Lett. {\bf B440} (1998) 246.
\bibitem{Ish93} T. lshii \emph{et al.,} Z. Phys.  {\bf A347} (1993) 41.

\end{thebibliography}
\end{document}